\documentclass[universe,article,accept,pdftex,moreauthors]{Definitions/mdpi}

\firstpage{1} 
\pubvolume{1}
\issuenum{1}
\articlenumber{0}
\pubyear{2026}
\copyrightyear{2026}
\externaleditor{Academic Editor: Firstname Lastname}
\datereceived{21 May 2026} 
\daterevised{21 June 2026} % Comment out if no revised date
\dateaccepted{26 June 2026} 
\datepublished{ } 
\hreflink{https://doi.org/} % If needed use \linebreak
\usepackage{pifont}
\usepackage{graphicx}
\usepackage{natbib}
\usepackage[utf8]{inputenc}

\DeclareUnicodeCharacter{200B}{}  % Ignora zero-width space
\providecommand{\aj}{AJ}

\usepackage{etoolbox}
\pretocmd{\bibliography}{\catcode`\&=12 }{}{}
\Title{The Limits of Photometric Dynamics: Benchmarking Cluster Relaxation Diagnostics}

\Author{Alisson P. Costa
$^{1, 2,}$*
$^{,\dagger,\ddagger}$, Andre L. 
 B. Ribeiro $^{2,\ddagger}$,  Zhonglue L. Wen 
 $^{3,\ddagger}$ and
Flavio R. Morais-Neto $^{4,\ddagger}$} 

\AuthorNames{Alisson Costa, Andre Ribeiro, Zhonglue Wen and Flavio Morais-Neto}

\address{%
$^{1}$ \quad Programa de Pós-Graduação em Modelagem Computacional, Universidade Estadual de Santa Cruz, \mbox{Ilh\'eus 454650-000, BA, Brazil} \\

$^{2}$ \quad Laboratorio de Astrofisica Teorica e Observacional, Universidade Estadual de Santa Cruz, \mbox{Ilh\'eus 454650-000, BA, Brazil}
\\ 

$^{3}$ \quad National Astronomical Observatories, Chinese Academy of Sciences, 20A Datun Road, Chaoyang District, Beijing 100012, China\\
$^{4}$ \quad Instituto de Astronomia, Geociências e Ciências Atmosféricas, Universidade de São Paulo, \mbox{São Paulo 05508-090, SP, Brazil} \ 
} 

\corres{Correspondence: apcosta.ppgmc@uesc.br }

\firstnote{Current address: Affiliation
.}  
\secondnote{These authors contributed equally to this work.
}

\abstract{
Galaxy clusters are key probes of cosmology and structure formation, yet their dynamical classification traditionally relies on spectroscopic redshifts, which do not scale efficiently with survey size. As large photometric surveys such as LSST become available, photometric redshifts offer an attractive alternative, but their impact on velocity-based diagnostics remains poorly constrained. 
{We quantify the sensitivity of two Gaussianity diagnostics---the Anderson--Darling (AD) test and Gaussian mixture modeling (Mclust)---to different photometric redshift error prescriptions. By propagating Gaussian and Student-\emph{t} uncertainties through velocity distributions constructed from SDSS photometric redshifts, we assess how the choice of error model affects the recovery of cluster dynamical states established by the independent \(\Gamma\) morphological proxy.}
Using 1672 SDSS clusters with pre-existing relaxation parameters ($\Gamma$), we perform Monte Carlo resampling under Gaussian and Student-$t$ error models, the latter used to mimic heavy-tailed uncertainties and catastrophic outliers. We also conduct a spectroscopic control experiment in which mock photometric redshifts are generated from spectroscopic measurements.
Under Gaussian errors, relaxed clusters are recovered in $\sim$95\% of realizations, whereas unrelaxed systems are detected in only $\sim$5\%, revealing a strong bias toward relaxed classifications. Student-\(t\) errors reduce relaxed recovery to $\sim$60--$70\%$ and increase unrelaxed recovery to $\sim$30--$45\%$, although this remains incomplete. Paired Wilcoxon tests confirm that these differences are statistically significant. 
This limitation has direct implications for large photometric surveys, suggesting that dynamical studies based primarily on photometric data may significantly underestimate the fraction of disturbed clusters unless supported by robust spectroscopic calibration, catastrophic-outlier mitigation, and validation with realistic survey mock catalogs.}


\begin{document}
%%%%%%%%%%%%%%%%%%%%%%%%%%%%%%%%%%%%%%%%%%

\section{Introduction}

Since Edwin Hubble’s observations in 1923--1924 established that spiral “nebulae” were in fact galaxies external to the Milky Way, the~study of galaxy clusters---recognized as some of the largest gravitationally bound systems in the Universe---has progressed substantially. Over~time, clusters have become essential tools both as cosmological probes and as laboratories for studying galaxy evolution and the physical processes governing their environments~\cite{2005ApJ...623..721P, 2017AJ....154...96D,
2024MNRAS.535.1348C}.

In this context, a~fundamental aspect of the cluster environment is the identification of substructures, commonly associated with galaxy groups in which the cluster gravitational potential has not yet fully virialized the constituent galaxies (e.g., \cite{2019arXiv190109198F, benavides2023dsp}). To~identify such systems, one of the least computationally demanding approaches consists of assessing deviations of the cluster velocity distribution from a normal distribution with similar statistical properties~\cite{1993AJ....105.1596B, Ribeiro2013}.

This principle motivates the application of statistical normality tests to the line-of-sight velocity distribution, such as the Anderson–Darling (AD) test (e.g.,  \cite{2012MNRAS.421.3594H}), as~well as the identification of multiple kinematic components through Gaussian mixture models (e.g.,~\cite{2015A&A...580A..69E}). However, these methods are typically applied only to systems with spectroscopic information for their member galaxies, since spectroscopic redshifts---forming the basis of all other kinematic properties---provide the lowest measurement uncertainties (e.g.,~\cite{1998ApJ...492...98G}). On~the other hand, when applied to photometric data, these techniques may lead to biased or unreliable results due to the substantial uncertainties associated with photometric redshifts (also referred to as photo-\(z\)) (e.g., \cite{2010ApJ...725..794Q}).

{The origin of these uncertainties lies in the way photometric redshifts are estimated. Photometric redshifts are distance estimates derived from the observed fluxes of galaxies in a finite set of broad photometric bands. In~practice, the~measured colors are compared either with spectral-energy-distribution (SED) templates shifted over a discrete grid of trial redshifts, or~with empirical and machine learning relations calibrated on galaxies with known spectroscopic redshifts (e.g., \cite{2024MNRAS.530.2012T, 2026JApA...47...28F}). Because~broad-band photometry samples the SED much more coarsely than spectroscopy, photometric redshifts are available for vastly larger galaxy samples, but~they come with substantially larger uncertainties and are more prone to color--redshift degeneracies and catastrophic outliers.}

While it is widely assumed that photometric redshift uncertainties degrade dynamical diagnostics, a~quantitative and systematic assessment of this effect remains lacking.
Particularly, in~the current era of increasingly large data volumes, where even minimally sophisticated computational approaches incur substantial computational costs~\cite{2018arXiv181111880J}, techniques such as those previously mentioned---that use few parameters---can become important starting points for more detailed dynamic diagnostics a~posteriori.

For example, the~Legacy Survey of Space and Time (LSST;~\cite{lsstsciencecollaboration2009lsstsciencebookversion, 2019ApJ...873..111I}), which is being conducted by the Vera C. Rubin Observatory, will survey the entire sky accessible from its location in the northern Chilean desert, combining survey area and depth in a manner unprecedented among optical surveys. Initial estimates indicate that over the 10 years of survey operations, approximately 20 TB of data will be produced per night, resulting in a total data volume of about 15 PB, equivalent to 20 billion galaxies 
({\url{https://rubinobservatory.org/for-scientists/rubin-101/key-numbers} (accessed on 27 April 2026)). In this regime, even small methodological biases can propagate to billions of systems, potentially leading to systematic misclassification of cluster dynamical~states.

Thus, the wide availability of photometric redshifts from large surveys motivates a critical question: \emph{how sensitive are traditional velocity-based dynamical diagnostics to the uncertainties inherent in photometric redshifts?} Rather than testing the methods on raw photometric data---which would require a complete simulation of the photometric redshift estimation pipeline---we adopt a controlled approach: we quantify how different prescriptions for photometric redshift errors (Gaussian versus heavy-tailed) affect the recovery of dynamical classifications established by an independent morphological proxy. Considering the extensive study by Ref. \cite{2017AJ....154...96D}, it is expected that these methods will achieve an accuracy rate of at least 70\% for the evaluated systems (even in photometric data), in order to obtain a significant difference in the other properties between the systems. Achieving this level of performance indicates that, although photometric redshifts inevitably carry significant uncertainties, their massive volume could represent an essential resource for preliminary insights into the dynamics of clusters in the era of LSST.

Given this context, we investigated the (in)ability and possible reliability of classical one-dimensional Gaussianity tests when applied to photometric redshifts. We apply these tests to a real observational dataset with a pre-established dynamical state, allowing us to directly compare their output against this known benchmark. Our goal is not to replace spectroscopic analyses or invalidate more detailed dynamic analyses on photometric data, but~to quantify the level of degradation of statistical tests commonly applied to spectroscopic data. In~other words, whether they can (or, once and for all, cannot) serve as initial and scalable diagnostics when applied to data containing only photometry. Throughout the paper, we assume a flat $\Lambda$CDM model with $\Omega_m = 0.3$, $\Omega_\Lambda = 0.7$, $H_0 = 70 \, \mathrm{km/s/Mpc}$.

%%%%%%%%%%%%%%%%%%%%%%%%%%%%%%%%%%%%%%%%%%
\section{Data and~Methodology}
\vspace{0.3cm}
\subsection{Background}

It is well known that after the relaxation process that occurs in galaxy clusters, the~three-dimensional velocity distribution of galaxies is well described by a Maxwell--Boltzmann distribution, characteristic of a system that has reached isotropy and equilibrium (e.g., \cite{1967MNRAS.136..101L}). When this velocity distribution is projected onto the line of sight, a~Gaussian distribution is obtained as a result~\cite{beers1991dynamical, Hansen_2006, hou2009statistical}.

In this sense, when the velocity distribution of a galaxy cluster deviates significantly from a Gaussian, its internal dynamics are likely out of equilibrium. Such deviations often manifest as multimodal patterns or other non-Gaussian features (e.g., skewness or heavy tails), typically caused by recent or ongoing merger of two comparably massive groups or the accretion of a smaller group by a big host, as~studied by~\cite{2012A&A...540A.123E, Ribeiro2013, 2017AJ....154...96D}. Generally, the~actors involved during and after this process are the so-called~substructures.

Consequently, the~cluster sample used in this work is drawn from~\cite{2013MNRAS.436..275W}, whose dynamical states have been previously calibrated and validated using the photometrically derived $\Gamma$ indicator. This parameter was optimized and tested against a reference sample of clusters with known dynamical states---classified as relaxed or unrelaxed based on X-ray, spectroscopic, and~radio data---achieving a $94\%$ success rate in identifying disturbed systems. Accounting for the uncertainties inherent to photometric redshifts, we assess the agreement between $\Gamma$ values and line-of-sight velocity distributions. Using standard spectroscopic indicators, we quantify the fraction of systems classified as substructured by $\Gamma$ that exhibit non-Gaussian features, as~well as the fraction of relaxed systems that retain a Gaussian~signature.

\subsection{Data}
\label{sec22}

Our analysis is based on the 2092 rich clusters identified by~\cite{2013MNRAS.436..275W} from SDSS-DR8~\cite{2011ApJS..193...29A} data, selected within the redshift range $0.05 \leq z < 0.42$ and with mass $M_{200} > 1.0 \times 10^{14} M_{\odot}$, where the parent catalog achieves a completeness exceeding $95\%$ (see~\cite{2012ApJS..199...34W}).
Their catalog implements the ref.~\cite{2009ApJS..183..197W} methodology to deliver, for~each cluster, the~luminous member galaxies, a~richness estimate, and~the radius within which the dynamical state is quantified via the \(\Gamma\) indicator.

The location of the first brightest cluster galaxy (BCG) identified from optical photometric data is adopted as the cluster center. This choice is justified because, for~clusters with X-ray emission,~Ref. \cite{2009ApJS..183..197W} demonstrated that most primary BCGs have a projected separation of less than $0.2$ Mpc from the X-ray~peaks.

Candidate luminous member galaxies for a cluster are chosen based on their redshifts and how far they are from the center of the cluster. The~uncertainties in the photometric redshift \(\sigma_z\) are approximately in the range of \([0.025-0.030]\) for redshifts smaller than $0.45$. {This uncertainty is related to what is determined spectroscopically \cite{2012ApJS..199...34W}}. Ref. \cite{2009ApJS..183..197W} also assumes that the uncertainties of the redshifts for all galaxies scale with the relation \(\sigma_z = \sigma_0 (1+z)\). Member galaxies with extinction-corrected magnitude \(M_r^{e}(z) \leq -20.5\) are selected within a photometric redshift range of \(z\pm 0.04(1+z)\). The~\(M_r^{e}(z)\) is obtained by \(M_r^{e}(z) = M_r(z) + Qz\), with~$Q = 1.62$ \cite{2003ApJ...592..819B}, so that for 2092 rich clusters, this selection has a completeness greater than $90\%$ \cite{2009ApJS..183..197W}. Objects with high photometric error, $zErr > 0.08(1+z)$, were discarded due to probable low photometric quality or possible stellar contamination. Despite the care taken in the selection, this redshift range can introduce contamination by foreground and background galaxies. To~mitigate this effect, the~photometric redshift selection was refined using spectroscopic data from SDSS-DR9~\cite{2012ApJS..203...21A}, so that galaxies outside the photometric list, but~with spectroscopic data, are included as members if they meet the separation (<2$ \,\,\mathrm{Mpc})$ and velocity $(\Delta v \leq 2500 \,\,\mathrm{km \cdot s^{-1}})$ criteria from the cluster center \cite{2013MNRAS.436..275W}.

Following~\cite{2012ApJS..199...34W}, the~cluster richness and radius \(r_{200}\) are derived from this refined member list. For~each cluster, the~total r-band luminosity within 1 Mpc \((L_{1\,Mpc})\) is calculated, after~background subtraction. \(L_{1\,\mathrm{Mpc}}\) is then converted to \(r_{200}\)---the radius where the mean density is 200 times the critical density---via the scaling relation from~\cite{2012ApJS..199...34W} given by
\begin{equation}
\log r_{200} = -0.57 + 0.44 \log(L_{1\,\mathrm{Mpc}} / L^*).
\label{r200_luminosity}
\end{equation}

Here, $r_{200}$ is in Mpc and $L^{*}$ denotes the evolved characteristic luminosity in the r-band, calculated from \(L^{*}(z) = L^{*}(z=0)10^{0.4Qz}\) \cite{2003ApJ...592..819B}. The~total
r-band luminosity $L_{200}$ is computed by summing the luminosities of member galaxies within $r_{200}$, and~richness is defined as $R_{L^{*}}=L_{200}/L^{*}$. 

The choice of this sample is motivated by the availability of an independent dynamical proxy, the~$\Gamma$ parameter (briefly described in the next section), which is derived from the projected spatial distribution of galaxies rather than from their velocity information. This allows a direct and non-circular comparison between structural and kinematic diagnostics. The~large size of the sample further ensures statistical robustness, allowing us to identify systematic effects rather than fluctuations driven by individual systems. Moreover, the~use of observational data ensures that the analysis incorporates realistic measurement uncertainties and selection effects, which are often difficult to capture in idealized simulations. This makes the sample particularly suitable for our purpose, as~it enables a controlled assessment of how photometric uncertainties affect the performance of classical dynamical diagnostics.
For more details on the sample clusters and the selection criteria used, see~\cite{2012ApJS..199...34W,2009ApJS..183..197W}.

\subsection{Relaxation Parameter \(\Gamma\) for Galaxy~Clusters}
\label{gamma}

The dynamical state of the 2092 rich clusters was previously quantified by~\cite{2013MNRAS.436..275W} using two-dimensional maps of the brightness distributions of the member galaxies of each studied field.
The spatial distribution of member galaxies, selected as described earlier, is smoothed with a Gaussian kernel weighted by each galaxy's luminosity. This produces a projected brightness distribution where light traces~mass.

To characterize dynamical state, Ref.~\cite{2013MNRAS.436..275W} calculates three metrics within the central region ($r_{500}=(2/3)r_{200}$, following~\cite{2003ApJ...590..197S}) where member contamination is lower:

\begin{itemize}
    \item The asymmetry factor (\(\alpha\));
    \item The ridge flatness (\(\beta\)); 
    \item The normalized deviation (\(\delta\)). 
\end{itemize}

In short, we will describe the main steps for obtaining the relaxation parameter \(\Gamma\). Initially, a~smoothing of the optical maps of each cluster is performed, where the coordinates of the galaxies (Ra and Dec) are converted into Cartesian coordinates, so that the optical luminosity within each pixel \((x_i , y_j )\) can be calculated through the convolution of all members with 
the following Gaussian kernel:
\begin{equation}
I(x_i, x_j) = \sum_{k=1}^{N_{200}} L_k \, g(x_i - x_k, y_j - y_k, \sigma_k),
\label{smoothed_luminosity}
\end{equation}
\noindent{where $x_k$ and $y_k$ are the coordinates of the $k$th member galaxy, $L_k$ its $r$-band luminosity in units of $L^*$, $N_{200}$ the number of members within $r_{200}$ and a photometric redshift slice of $z \pm 0.04(1+z)$, and~$g(x,y,\sigma)$ a 2D Gaussian with smoothing scaled by $\sigma$.}

After this step, the~metrics that yield the \(\Gamma\) parameter are obtained through the following equations:
\begin{equation}
\alpha = \frac{\Delta^2}{S^2}, \quad
\beta = \frac{c_{200}^R}{\langle c_{200} \rangle}, \quad
\delta = \frac{\sum_{i,j} \left[ I(x_i, y_j) - I_{\mathrm{2Dmodel}}(x_i, y_j) \right]^2}{S^2}.
\label{all_param}
\end{equation}

{These three metrics define a three-dimensional parameter space, in~which relaxed and unrelaxed clusters are separated by a boundary plane. The~distance of a cluster to this plane is given by the parameter \( \Gamma \).}

With respect to the asymmetry factor (\(\alpha\)), \(S^2 = \sum_{i,j} I^2(x_i, y_j)\) represents the total influence of internal fluctuations within \(r_{500}\), and~$\Delta^2 = \frac{1}{2} \sum_{i,j} \left[ I(x_i, y_j) - I(-x_i, -y_j) \right]^2$ the difference in these fluctuations for symmetrical pixels. In~other words, \(\alpha\) is related to the regularity of morphological symmetry, commonly exhibited in clusters considered relaxed (without substructures). \(\beta\) (ridge flatness) measures the brightness profile of the studied region, considering the fact that in a relaxed cluster, the~smoothed optical map generally presents a pronounced surface brightness profile in all directions, and Ref.~\cite{2013MNRAS.436..275W}  obtains the light profile in various directions and fits the profile with King's one-dimensional model \cite{1962AJ.....67..471K}. From~this, for~each direction around the cluster, they calculate \(c_{200} = r_{200}/r_0\) (with \(r_0\) via King's model), obtaining \(c_{200}^{R}\)---the minimum value among the calculated \(c_{200}\) values, indicating a more diffuse profile (possibly due to mergers or substructure infall)---and the mean of the calculated concentration factors, \(\langle c_{200} \rangle\). Thus, for~a relaxed system, \(\beta \sim 1\) is expected, while for non-relaxed systems, the~flatter ridge in the map gives a smaller \(\beta\).

For the normalized deviation, \(\delta\), the~smoothed map of the cluster is first modeled using the two-dimensional elliptical King model ({$r_{\mathrm{iso}}$ is the cluster-centric `distance' of an isophote given by $r_{\mathrm{iso}}^2 = \left[ x \cos(\theta) + y \sin(\theta) \right]^2 + \epsilon^2 \left[ -x \sin(\theta) + y \cos(\theta) \right]^2$, where $\epsilon$ denotes the axial ratio (semiminor over semimajor axis) of the isophote ($\epsilon \leq 1$), and~$\theta$ is the position angle of its major axis.}) (\(I_{\mathrm{2Dmodel}}(x, y) = I_0/[1 + (r_{\mathrm{iso}} / r_0)^2]\)), allowing the creation of a model of how the system would be if it were relaxed. Thus, the~larger the \(delta\), the~greater the complexity not captured by the model, suggesting that the cluster is unrelaxed. Therefore, \(\delta\) quantifies the residual complexity of the maps, so the higher the \(\delta\) value, the~more asymmetries, substructures, or~irregularities there are in the~cluster.

Using these metrics, the~separation plane shown in Figure~\ref{planogama} is calculated by \mbox{\(\beta=A\,\alpha+B\,\delta+C\)}, with~the optimal plane defined for $A = 1.9$, $B = 3.58$, and~$C = 0.1$. Finally, the~relaxation parameter that quantifies the dynamic state of the clusters is calculated by
\begin{equation}
\Gamma =[\beta-(1.90\,\alpha+3.58\,\delta+0.10)]/k,
\label{gamma}
\end{equation}

\noindent{so that $k=\sqrt{1+A^2+B^2}$ and in practical terms the constant factor is assumed to be $k=1$ by~\cite{2013MNRAS.436..275W}. Thus, the~\(\Gamma\) parameter is
positive for relaxed clusters and negative for unrelaxed clusters. A~higher value of \(\Gamma\) indicates that the cluster is more consistent with a relaxed state. 
Although \(\Gamma\) is derived from photometric data, it is based on spatial morphology rather than velocity information, making it an independent proxy for dynamical state when compared to velocity-based diagnostics. For~more technical details on this methodology, see~\cite{2013MNRAS.436..275W}. }

\begin{figure}[H]
 \centering
\includegraphics[width=0.6\linewidth]{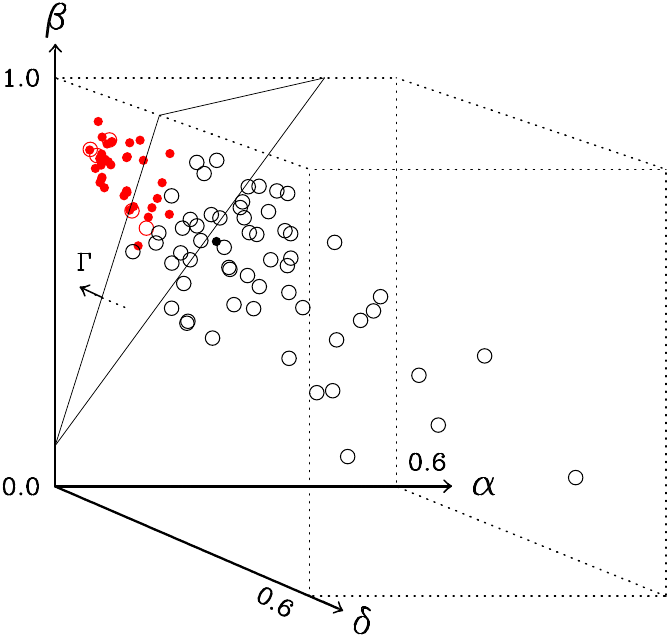}
    \caption{Figure 4 from \cite{2013MNRAS.436..275W} shows the distribution of relaxed (red circles) and unrelaxed (open circles) clusters
 in the three-dimensional parameter space defined by \(\alpha\), \(\beta\) and \(\delta\) used to validate the method. The~two populations are well separated by the plane indicated with thin solid lines. The~dynamical state of a cluster is quantified by \(\Gamma\), defined as its distance to this plane. Solid black dots and open red circles - discussed in Appendix A by \cite{2013MNRAS.436..275W} - represent systems with \(\Gamma\) parameter inconsistent with the dynamics established by previous studies.
{Image reprinted from \mbox{Ref.~\cite{2013MNRAS.436..275W}}, by~permission of Oxford University Press. This image is not covered by the terms of the Creative Commons license of this publication. For~permission to reuse, please contact the rights holder.}}
    \label{planogama}
\end{figure}

{We emphasize that the use of \(\Gamma\) does not imply that photometric data alone can replace spectroscopy. A~robust characterization of dynamical state ultimately requires both spatial and kinematic information. The~\(\Gamma\) parameter is employed here strictly as an independent photometric proxy, derived from the projected optical luminosity distribution, and~therefore does not rely on any assumption about the velocity distribution of cluster members. This independence allows us to test how the photo-\(z\)-based estimators, AD and Mclust, respond to different photometric uncertainty models, and~whether they yield consistent classifications---without circularity. We acknowledge the limitations of photometric redshifts for individual galaxy velocities; however, given the limitations imposed by spectroscopy,  the~\(\Gamma\) parameter offers a statistically valid and computationally efficient way to probe the dynamical state of a large population with this limitation.}

%%%%%%%%%%%%%%%%%%%%%%%%%%%%%%%%%%%%%%%%%%%%%%%%%%%%%%%%%%%%%%%%%%%%%%%%%%%%%%%%%%%%%%%%%%%%%%%%%%%%%%%%%%%%%%%%%%%%%%%%%%%%%%

\subsection{Probing Velocity Distributions Using Photometric~Redshifts}
\label{probing}

{The application of dynamical diagnostics---which are routinely used with spectroscopic redshifts---to photometric data is fundamentally challenged by the substantial uncertainties in photometric redshift estimates. A~typical photometric redshift error of  $\sim$0.02(1 + \emph{z}) (e.g., \cite{2006A&A...457..841I}) corresponds to $\sim$6000 km/s---far greater than the intrinsic velocity dispersion of galaxy clusters, which lies between $500$ and $1200$ km/s~\cite{Ferragamo2021Velocity, Girardi1993Velocity}. These large uncertainties can erase or distort dynamical signatures in the velocity distribution. However, a~systematic quantification of how different error models affect the recovery of dynamical classifications remains lacking. In~this work, we do not attempt to reproduce a full photometric redshift estimation pipeline. Instead, we adopt a controlled Monte Carlo approach to quantify the sensitivity of the AD and Mclust diagnostics to the assumed photometric error distribution. This allows us to isolate the effect of the error model from other sources of uncertainty, providing a baseline for interpreting results from future photometric surveys.}

{Therefore, before~describing the sample selection and the Monte Carlo propagation of photometric redshift uncertainties, we first present the statistical diagnostics used in this work. In~the following subsections, we first describe the assumptions, outputs, and~physical meaning of these statistical tools, and~then present the cluster selection, the~photometric redshift resampling procedure, and~the definition of the recovery fractions.}

\subsubsection{Statistical Diagnostics: Assumptions and~Interpretation}

{In this work, the~dynamical information contained in photometric velocity distributions is evaluated through two  statistical diagnostics: the Anderson--Darling (AD) normality test and Gaussian mixture modeling through Mclust. We provide here a more explicit description of their assumptions and of how their outputs are interpreted in the context of cluster dynamical classification.}

{The Anderson--Darling (AD) test is an empirical-distribution-function test designed to quantify the discrepancy between an observed sample and a reference cumulative distribution \cite{stephens1974edf}. In~the present application, the~null hypothesis is that the line-of-sight velocity distribution of a given cluster is compatible with a Gaussian distribution. This choice is physically motivated by the expectation that dynamically relaxed systems have approximately Gaussian projected velocity distributions, whereas dynamically disturbed systems may exhibit departures from Gaussianity, such as skewness, heavy tails, or~multimodal structure (e.g., \cite{hou2009statistical, 2018MNRAS.475.4704R}).
In this test we adopt a significance level of \(\alpha=0.05\): when \(p\)-value \(>\alpha\), Gaussianity is not rejected and the realization is considered compatible with a relaxed velocity distribution; when \(p\)-value \(\leq\alpha\), Gaussianity is rejected and the realization is considered non-Gaussian.}

{The use of the AD test is also justified because it is straightforward to apply to unbinned data and does not introduce the binning sensitivity highlighted by~\cite{2019JCAP...02..005P} for the Kolmogorov--Smirnov case, for~example. Although~the AD test belongs to the Cramér--von Mises class of EDF tests (e.g., \cite{10.1214/25-EJS2386}), for~which binning effects have been suggested as a potential concern in discretised applications, we apply it directly to the continuous line-of-sight velocity measurements, following standard practice in the literature (e.g., \cite{2023A&A...676A.127D,2025A&A...699A..88P}). Thus, even if binning sensitivity studies existed for the AD test in categorical contexts, they would not be directly relevant to our analysis.}

{The second diagnostic is based on Gaussian mixture modeling, implemented with Mclust \cite{fraley2007model}. This approach assumes that the observed velocity distribution can be represented as a mixture of Gaussian components. The~model parameters are estimated through the Expectation--Maximization (EM) algorithm~\cite{Scrucca2023Model-Based}, and~competing models are compared using the Bayesian Information Criterion (BIC). In~this study, we restrict the comparison to two physically interpretable cases: \(G=1\), corresponding to a single Gaussian component, and~\(G=2\), corresponding to two Gaussian components. A~model with \(G=1\) is interpreted as a simple velocity distribution, consistent with a relaxed system, whereas \(G=2\) indicates that the distribution is better described by two kinematic components, suggesting possible substructure or dynamical complexity.}

{Because photometric redshift uncertainties are large compared to the intrinsic velocity dispersion of galaxy clusters, a~single application of either diagnostic would not provide a robust classification. We therefore use a Monte Carlo approach, best described after the description of the sample of galaxy clusters used below.}

\subsubsection{Sample Selection and Photometric Redshift~Resampling}

To access the velocity distributions of the photometric cluster sample, we initially applied some restrictions to the sample in order to avoid systems without well-resolved photometric characteristics, i.e.,~without significant distance from the separation plane. To~do this, we calculated for each cluster the significance ratio $\xi = \Gamma/\sigma_{\Gamma}$, which measures how many individual standard deviations $\Gamma$ is away from the physical classification threshold $\Gamma = 0$. From~this, we applied a significance threshold $|\Gamma/\sigma_{\Gamma}| \geq 2$ to select only clusters with statistically robust dynamical classification. This criterion, equivalent to $2\sigma$ under Gaussian errors, guarantees confidence $\geq 97.5\%$ in each classification, clearly separating the relaxed ($\Gamma/\sigma_{\Gamma} \geq 2$) and unrelaxed ($\Gamma/\sigma_{\Gamma} \leq -2$) states from the intermediate region ($|\Gamma/\sigma_{\Gamma}| < 2$). Figure~\ref{zones} represents this~procedure.

\begin{figure}[H]
	
	\includegraphics[width=0.8\linewidth]{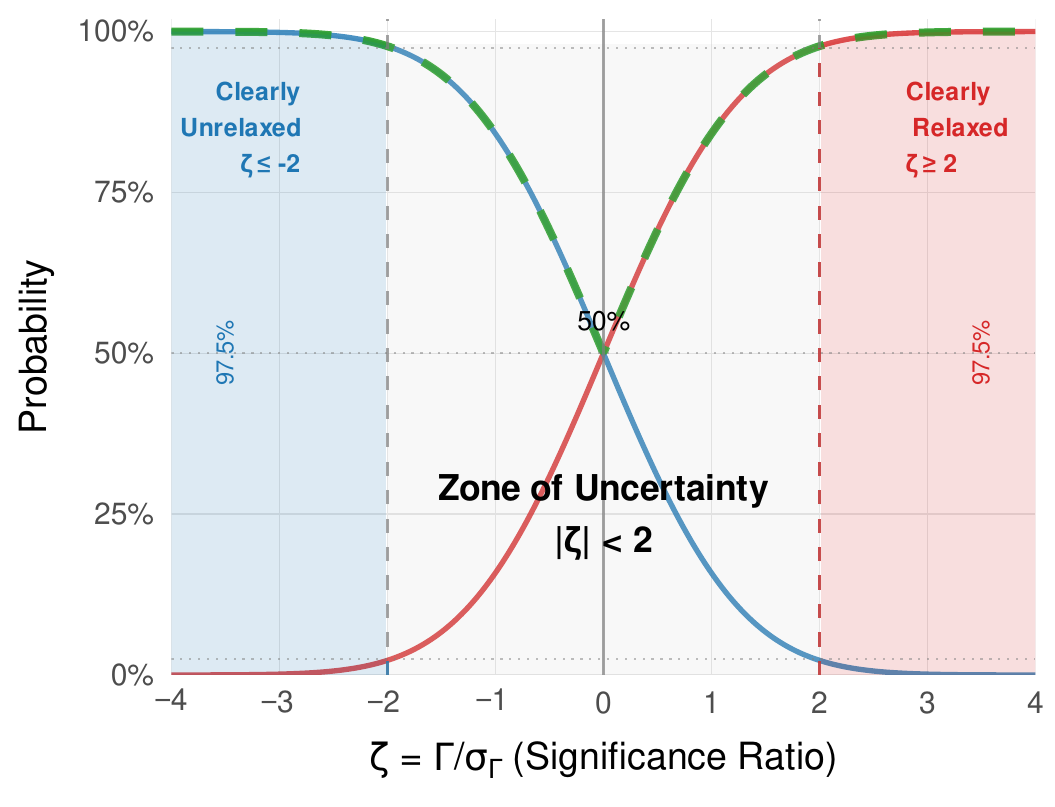}
	\caption{Probability
 of relaxed/unrelaxed state as a function of the significance ratio for the cluster sample. The~red and blue curves indicate which state is most likely for each group as the \(\zeta\) values change. The~dashed green curve represents the reliability of the decision made for the systems based on the significance ratio. The~intersection of the probability curves defines the uncertainty region where there is a possibility of systems not being so well diagnosed photometrically via the \(\Gamma\) parameter, although~its success rate is \(94\%\) \protect~\cite{2013MNRAS.436..275W}.}
	\label{zones}
\end{figure}

{In addition to the previous restriction, we retained only clusters with at least 20 member galaxies within the region (\(r_{500}\)) where the parameter \(\Gamma\) was estimated. The~choice \(N \geq 20\) is not intended to imply that a system with exactly 20 members has a fully secure individual dynamical interpretation. Rather, it is a conservative minimum multiplicity threshold adopted to reduce severe sampling noise in velocity-distribution diagnostics and to ensure numerical stability of quantities such as velocity dispersion and Gaussianity tests. This threshold is consistent with previous work on group and cluster dynamics, which shows that statistical tests of velocity distributions and dispersion profiles become strongly sample-size-dependent and commonly adopt similar cuts (e.g., \(N \geq 20\)) when separating low- and high-multiplicity systems (e.g., \cite{Ribeiro2013, Lopes2023The, Brambila2023Examining}). Thus, the~\(N \geq 20\) criterion provides statistical stability at the ensemble level (here, 1672 clusters) and facilitates comparisons between richness sub-samples, while clusters with richness close to the threshold still require careful interpretation when analyzed individually. An~analysis of the cluster richness dependence will be presented in Section~\ref{sec34}.} After these considerations, with~a more restricted sample of 1672 clusters, we can carry out the following investigation:

\begin{itemize}
	\item[\ding{87}] To what extent can the dynamical information encoded in the \(\Gamma\) indicator be recovered through the use of photometric redshifts?
\end{itemize}

Due to the non-negligible nature of photometric errors, a~single measurement of the AD or Mclust test would obviously be inconclusive. To~try to account for this, we performed 1000 Monte Carlo simulations for each cluster. In~each run, galaxy photometric redshifts were resampled assuming Gaussian and Student-\(t\) error distributions. The~performance of the tests will be based on the fraction of runs in which the hypothesis being tested is recovered. To~this end, we separated the set of clusters into two groups: those considered relaxed (\(\Gamma >0\)) and those unrelaxed (\(\Gamma < 0\)).

At this point in the methodology, it is imperative to eliminate some statistical issues from the process. For~example, \textit{why model the error distribution using two distinct distributions for each dynamical group?
} The answer to this is that by performing this comparative sensitivity analysis between the dynamical groups established by \(\Gamma\), we intend to disaggregate---to the extent that the information limit allows---the effects of instrumental uncertainty from the effects of cluster physics, guided by the \(\Gamma\) relaxation parameter. In~this case, while Gaussian modeling assumes an ideal error behavior, Student-\(t\) modeling introduces heavy tails into the process, simulating the presence of catastrophic photometric outliers (e.g., \cite{Martin2024An}) common in surveys, without~requiring explicit modeling of their~origin.

{The Student-\(t\) model is therefore adopted as an alternative error prescription, not as an additional dynamical test. Compared with a Gaussian distribution with the same central scale, a~Student-\(t\) distribution with \(\nu=3\) assigns a larger probability to large deviations from the mean. In~practice, most Monte Carlo realizations produce small redshift perturbations, as~in the Gaussian case, but~a non-negligible fraction produces much larger shifts. These rare but large perturbations mimic the effect of catastrophic photo-\(z\) outliers without requiring explicit identification of their physical origin for each galaxy. Thus, for~each statistical test considered (AD and Mclust), we compare the results obtained under the two error prescriptions: Gaussian errors mainly broaden and smooth the reconstructed velocity distribution, while Student-\(t\) errors occasionally generate extended tails and outlier-like distortions.}

Through this procedure, we hope to quantify the degradation of the AD and Mclust tests. We chose not to use a fully Bayesian framework, as~our goal is not to recover the intrinsic velocity distribution, but~to quantify the degradation of classical diagnostics under realistic observational noise, and~a fully Bayesian framework would partially absorb photometric uncertainties into the model, potentially masking the intrinsic limitations of the diagnostics themselves. Furthermore, the~computational cost of hierarchical Bayesian approaches should make them impractical for large-scale surveys such as LSST, where fast and scalable diagnostics are required. By~contrast, our approach deliberately isolates these~limitations.

Our approach can be expressed in mathematical form for Gaussian errors as follows:
\begin{equation}
	z_{\text{MC}, i}^{(j)} = z_{\text{photo}, i} + \delta_{i}^{(j)}; \,\,\,\text{with}\,\,\, \delta_{i}^{(j)} \sim \mathcal{N}(0, \sigma_{z, i}^2),
	\label{z_mc_gaussian}
\end{equation}

\noindent where \(\sigma_{z,i}\) represents, henceforth, the~formal errors from the SDSS photometric redshift associated with the \(z_{\text{photo}}\) of each \(i\)-th galaxy, already discussed in Section \ref{sec22}. The~\(\delta_{i}^{(j)}\) is the measurement error, assumed to be Gaussian, added to the initial~observation.

For the Student-\(t\) model, we consider that the photometric information for each Monte Carlo follows the distribution
\begin{equation}
	z_{\text{MC}, i}^{(j)} \sim t_{\nu}\left( \mu = z_{\text{obs}, i}, \, \text{scale} = \sigma_{z, i} \right),
	\label{z_mc_student}
\end{equation}

\noindent with degrees of freedom fixed at \(\nu=3\), representing a conservative choice between Gaussian noise and the presence of catastrophic outliers. This value has been adopted in astrophysical contexts to model both measurement noise and calibration uncertainties, providing robustness against outliers without requiring explicit outlier identification (e.g., \cite{2012ApJ...752...55K}). The~value of \(\mu\) represents the location parameter of the distribution, in~this case, fixed at the observed~redshift.

{For each Monte Carlo realization, the~result of the diagnostic is encoded in an indicator function. In~the case of the AD test, this indicator depends on the \(p\)-value associated with the null hypothesis for the evaluated dynamical group (relaxed and unrelaxed):}

\begin{itemize}
	\item Relaxed systems:
\begin{equation}
		\mathbb{I}^{(j)}_{\rm AD,relaxed} = 
		\begin{cases} 
			1, & \text{if } p^{(j)} > \alpha \quad (\text{H}_0 \,\,\,\text{not rejected}), \\
			0, & \text{if } p^{(j)} \leq \alpha \quad (\text{H}_0 \,\,\,\text{rejected}),
		\end{cases}
		\label{indicator_function_ad_relaxed}
	\end{equation}
	\item Unrelaxed systems:
\begin{equation}
		\mathbb{I}^{(j)}_{\rm AD,unrelaxed} = 
		\begin{cases} 
			1, & \text{if } p^{(j)} \leq \alpha \quad (\text{H}_0 \,\,\,\text{rejected}), \\
			0, & \text{if } p^{(j)} > \alpha \quad (\text{H}_0 \,\,\,\text{not rejected}).
		\end{cases}
		\label{indicator_function_ad_unrelaxed}
	\end{equation}
	
\end{itemize}

\indent As mentioned earlier, we use \(\alpha = 5\%\) as the significance level. For~the \(p\)-value calculation in the AD test, we used the \texttt{nortest} package in R (\cite{nortest}).

{By analogy with the AD test, the~Mclust indicator is defined according to the number of Gaussian components selected by the BIC:}

\begin{itemize}
	\item Relaxed systems:
\begin{equation}
		\mathbb{I}^{(j)}_{\rm Mclust,relaxed} = 
		\begin{cases} 
			1, & \text{if } G=1, \\
			0, & \text{if } G=2,
		\end{cases}
		\label{indicator_function_mclust_relaxed}
	\end{equation}
		
	\item Unrelaxed systems:
\begin{equation}
		\mathbb{I}^{(j)}_{\rm Mclust,unrelaxed} = 
		\begin{cases} 
			1, & \text{if } G=2, \\
			0, & \text{if } G=1.
		\end{cases}
		\label{indicator_function_mclust_unrelaxed}
	\end{equation}

\end{itemize}

The Mclust was applied by the \texttt{mclust} package also in R.
Since the choice of the best model for the velocity distribution is based on the one with the best BIC, we are actually asking the following question when applying Mclust: \textit{Is the velocity distribution of this cluster simple enough to be described by a single Gaussian (relaxed), or~is it complex enough to require at least two components (unrelaxed)?}

{Thus, in~general, \(\mathbb{I}^{(j)}=1\) always indicates that the expected dynamical state (according to the dynamic group evaluated) was recovered in the \(j\)-th Monte Carlo realization, while \(\mathbb{I}^{(j)}=0\) indicates that it was not recovered.} Finally, the~recovery fraction of tested hypotheses, for~example, of~relaxed systems (\(f_{\text{relaxed}}\)) in clusters with \(\Gamma >0\), is defined as the average over all \(N_{\text{MC}}\) Monte Carlo realizations:
\begin{equation}
	f_{\text{relaxed (or unrelaxed)}} = \frac{1}{N_{\text{MC}}} \sum_{j=1}^{N_{\text{MC}}} \mathbb{I}^{(j)}.
	\label{f_gauss}
\end{equation}

In other words, the~\(f_{\text{relaxed (or unrelaxed)}}\) indicator represents the robustness of dynamical identification in the face of photometric error. For~practical purposes, \(H_0\) corresponds to a Gaussian velocity distribution and, for~unrelaxed clusters, a~successful detection occurs when the test rejects the null hypothesis of~Gaussianity.

%%%%%%%%%%%%%%%%%%%%%%%%%%%%%%%%%%%%%%%%%%%%%%%%%%%%%%%%%%%%%%%%%%%%%%%%%%%%%%%%%%%%%%%%%%%%%%%%%%%%%%%%%%%%%%%%%%%%%%%%%%%%%%

\section{Results and~Analysis}

Considering the typical uncertainties of photometric redshift, there is a real risk that unrelaxed signatures may be masked by observational noise. In~this section, we aim to quantify the limitations by modeling the errors associated with photometric measurements using Gaussian and Student-t distributions. Taking this mistrust into account, we will also evaluate the recovery rate of dynamic states in order to establish the operational limits of the kinematic use of photometry compared to the structural robustness of the \(\Gamma\) indicator.

\subsection{On the Robustness of AD and Mclust Against Error Distributions in Relaxed (\(\Gamma > 0\)) and  Unrelaxed ($\Gamma < 0$) Clusters}
\label{sec31}

The results of the evaluation of the AD and Mclust methods on the relaxed sample are shown in Figure~\ref{gamma_relaxed}, where we observe a significant and systematic divergence in the fractions of recovered dynamic states when modeling the errors through a Gaussian and Student-\(t\) distribution. We chose to present the results using an overlay combining boxplots with violin plots, since this combination allows us to capture the analytical depth of the violin plots with the summary objectivity of the boxplots (e.g., \cite{Lan2021Visualization, Gerbing2024The, Marwan2025Recurrence}). From~Figure~\ref{gamma_relaxed}, while the Gaussian error model has a median of \(\sim0.95\) for the fraction of recovered relaxed diagnoses, exhibiting low dispersion, characterized by compact boxplots with short vertical lines (also called whiskers~\cite{Tanious2022Violin}), we see that when considering the Student-\(t\) model, we obtain a median of \(f_{relaxed}\) in the range of $\sim$0.60 to $0.70$, with~considerable~variability.

\begin{figure}[H]
	\centering
	\includegraphics[width=0.8\linewidth]{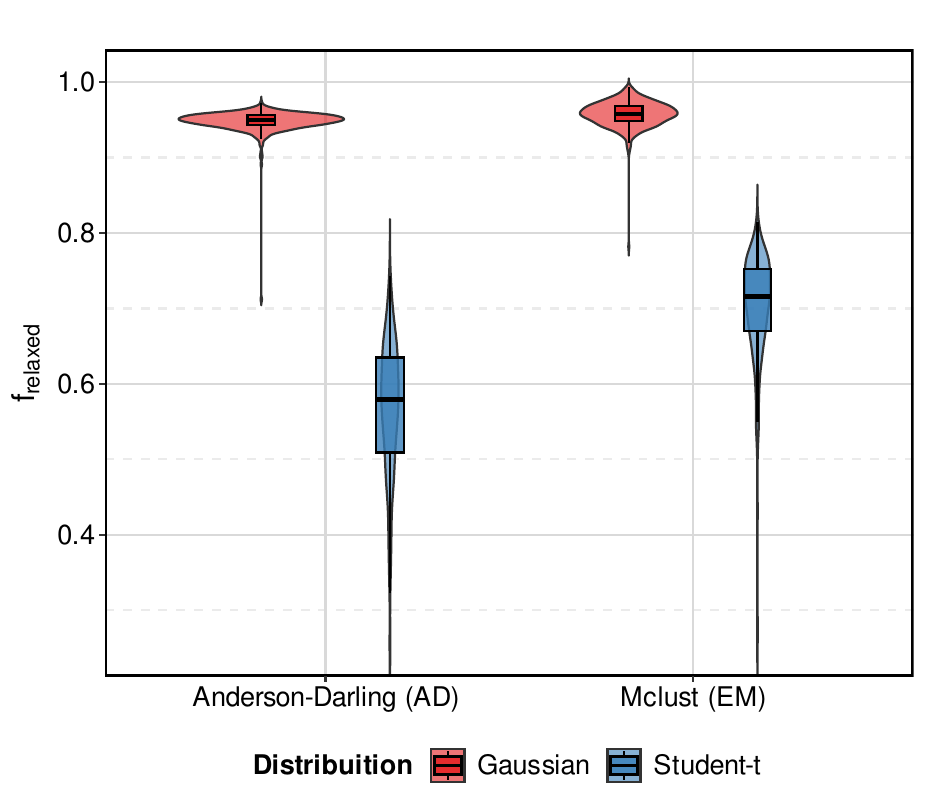}
	\caption{Sensitivity analysis of dynamic tests against different photometric error models. Under~Gaussian errors (red), both tests (AD and Mclust) show high stability ($f_{relaxed} > 0.9$). Under~Student-t errors (blue), which simulate catastrophic outliers, a~significant degradation in the ability to recover the dynamic status is observed, evidenced by the drop in medians and widening of~distributions.}
	\label{gamma_relaxed}
\end{figure}

{These results indicate that, for~the Gaussian error model (red symbols), the~photometric velocity distribution of relaxed clusters remains essentially Gaussian, as~the convolution with Gaussian uncertainties preserves the normality of the observed distribution}. Consequently, normality tests (AD and Mclust) rarely reject the null hypothesis. However, as~we will show below for unrelaxed clusters, this apparent stability comes at the cost of systematic contamination: the smoothing effect erases non-Gaussian signatures, causing intrinsically perturbed systems to be misclassified as Gaussian. On~the other hand, the~dispersion and significant drop in the median for models with Student-t error (blue symbols) reveals a possible degradation of dynamic information. The~introduction of heavy tails (simulating catastrophic outliers) distorts the shape of the velocity distribution enough so that, in~around \(30\%\) to   \(40\%\) of Monte Carlo simulations, the~clusters---although intrinsically relaxed---have inconclusive statistics or, depending on the threshold considered, they can be classified as unrelaxed, demonstrating the sensitivity of the methodology to the presence of photometric outliers common in surveys. This result, at~least preliminarily, indicates that, under~realistic Student-\(t\) errors, the~true success rate for identifying relaxed clusters drops to $\sim$60--70\%, compared to the $\sim$95\% suggested by Gaussian assumptions. Thus, the~degradation is not a failure of the Student-t model, but~should be seen as the exposure of the overconfidence of the Gaussian model~error.

\vspace{-3pt}

\begin{figure}[H]
	 \centering
	\includegraphics[width=0.8\linewidth]{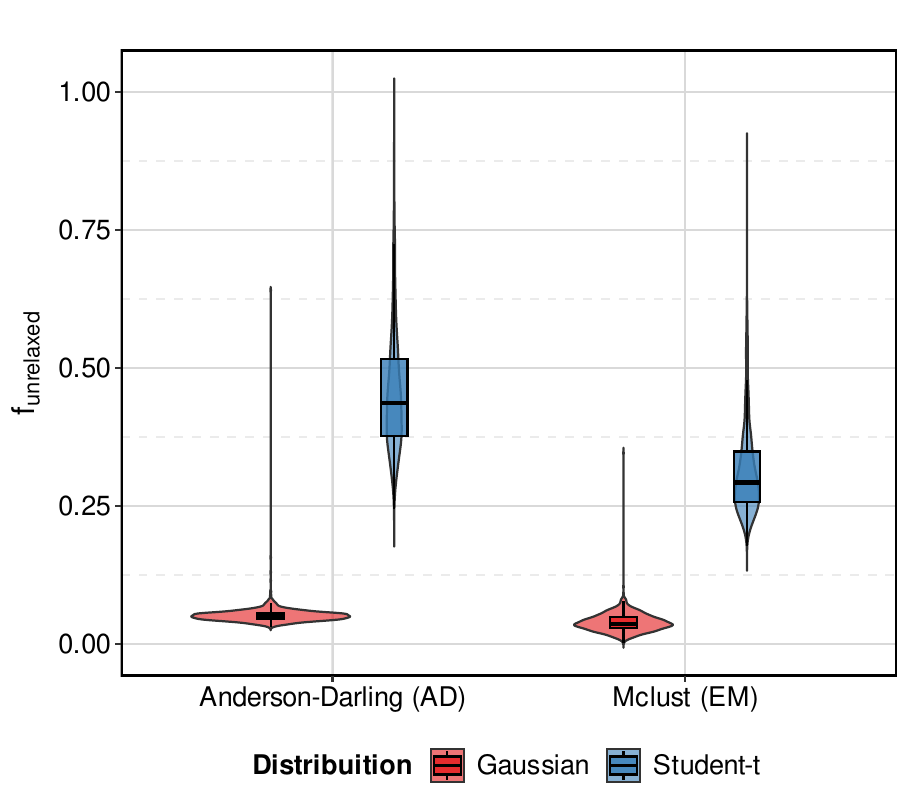}
	\caption{Sensitivity analysis for unrelaxed systems (similar to Figure~\ref{gamma_relaxed}) . Gaussian errors (red) result in negligible detection rates ($f_{unrelaxed} \sim 0.05$). In~contrast, adopting Student-t errors (blue) improves the recovery of unrelaxed signatures by accounting for heavy tails (catastrophic outliers). 
	}	
	\label{gamma_unrelaxed}
\end{figure}

At this stage, we evaluated the impairment of the dynamical classification yielded by the AD and Mclust tests in the sample of unrelaxed clusters. We seek the fractions of cases where the AD and Mclust tests classify the photometric velocity distributions as~non-Gaussian.

Although Figure~\ref{gamma_relaxed} shows that Student-\(t\) errors reduce the apparent success rates \(f_{relaxed}\) for relaxed clusters, it is questionable whether this is compensated for by better performance in unrelaxed systems. Figure~\ref{gamma_unrelaxed} directly tests this hypothesis. From~Figure~\ref{gamma_unrelaxed}, when analyzing first the Gaussian error model (idealized case)---superposition of violins and boxplots in red---we notice a significant flattening concentrated at the base of the figure, revealing a lack of sensitivity of the tests under this error modeling, where the underrepresentation of extreme velocities possibly  results in a failure to recover the unrelaxed nature of these clusters (predominance of false negatives). This compact shape indicates that the failure to diagnose the unrelaxed state was systematic across almost all simulations (evidenced by the boxplot compressed inside the violins, reflecting minimal variance). The~rare successful detections, represented by the whiskers, are likely attributable to stochastic sampling fluctuations inherent to finite datasets \cite{Armstrong2019Is, Papadopoulos2022The}, rather than a genuine sensitivity of the Gaussian error~model.

The Student-$t$ model raises $f_{\mathrm{unrelaxed}}$ from a negligible $\sim$5\% to a range of $\sim$30--45\%. While this represents a six-to-nine-fold improvement over the Gaussian baseline, it remains far from enabling reliable classification. This leads to a critical insight: neither model achieves robust identification of unrelaxed clusters, albeit for fundamentally different physical and statistical~reasons. 

The Gaussian error model fails by oversmoothing the velocity distribution, erasing multimodal features and leading to a failure to reject $H_0$ (Gaussianity) in $\sim 95\%$ of realizations, thereby systematically misclassifying perturbed systems as relaxed. In~contrast, the~Student-t model partially restores sensitivity to non-Gaussian features through its heavy tails, but~at the cost of increased dispersion and instability, which still prevents reliable~classification.

Photometric errors therefore modify the observed velocity distribution in two distinct ways: by smoothing it through Gaussian-like convolution or by introducing heavy-tailed distortions associated with catastrophic outliers. In~both cases, the~underlying dynamical information is degraded, though~via different~mechanisms.

\subsection{A Unified Assessment: Both Dynamical States Are~Compromised}
\label{sec32}

The comparison of Figures~\ref{gamma_relaxed} and \ref{gamma_unrelaxed} reveals a consistent pattern: when photometric errors are treated as Gaussian, both the AD and Mclust tests tend to classify clusters into a single dynamical state, regardless of their actual condition:

\begin{itemize}
	
\item {Relaxed clusters:} With Gaussian error assumptions, AD and Mclust identify Gaussian velocity distributions in $\sim$95\% (on average) of 
realizations per cluster. With~Student-\(t\) errors, this rate drops to $\sim$60--70\%, indicating that outlier contamination obscures the intrinsic Gaussianity (via photometric velocity distributions) of relaxed systems in roughly 30--40\% of realizations---suggesting that this type of contamination reduces the reliability of less rigorous identifications;
	
\item {Unrelaxed clusters:} Also under Gaussian error assumptions, AD and Mclust detect non-Gaussianity in only $\sim$5\% of realizations per cluster (median $f_{unrelaxed} \sim$0.05), meaning that 
intrinsically perturbed systems are overwhelmingly misclassified as relaxed (through null hypothesis rejection). When considering Student-\(t\) errors, detection improves to an average global recovery of
$\sim$30--45\% of realizations per cluster, but~the majority of realizations ($\sim$55--70\%) still fail to recover the true unrelaxed~state.	
	
\end{itemize}

In short, Gaussian error assumptions mask the true uncertainty in dynamical classification by artificially preserving Gaussianity in velocity distributions: for relaxed clusters this coincides with correct classification, but~for unrelaxed clusters it causes catastrophic misclassification as the intrinsic non-Gaussianity is smoothed~away.

The Student-t error model highlights that the performance of the tests is asymmetrically degraded under realistic photometric noise: relaxed states are moderately recoverable (60--70\%) and unrelaxed states remain largely undetected (only 30--45\% of the time).
Thus, the~Student-\(t\) model shows that even with a more realistic error model, the~classification scenario remains uneven and challenging when using velocity distributions derived from photometry. While relaxed systems maintain moderate identifiability under outlier contamination, unrelaxed systems remain largely obscured: Gaussian assumptions cause severe signal dilution, while even Student-\(t\) modeling only partially recovers the intrinsic non-Gaussianity, leaving the majority of perturbed clusters~misclassified.

Table~\ref{tab1} quantifies the results discussed for both samples in Figures~\ref{gamma_relaxed} and \ref{gamma_unrelaxed}.
However, to~assess whether these differences are systematic across the cluster sample, we applied paired Wilcoxon signed-rank tests (also known as ‘Mann--Whitney’ test) \cite{hollander2013nonparametric} to the cluster-by-cluster recovery fractions.
The Wilcoxon test is a non-parametric statistical hypothesis test used to compare paired samples. (e.g., \cite{2013arXiv1311.5354R,2016A&A...595A..73B}). The~statistical unit in this comparison is the cluster, not the individual Monte Carlo realization. For~each dynamical state and diagnostic, we compared the recovery fractions obtained under Gaussian and Student-\(t\) errors. The~results are reported in Table~\ref{wilcox}.

%%%%%%%%%%%%%%%%%%%%%%%%%%%%%%%%%%%%%%%%%%%%%%%%%%%%%%

\begin{table}[H]

	\caption{Mean and standard deviation of the recovery fractions for relaxed and unrelaxed clusters across different error models and~diagnostics.}
	\label{tab1}
	\begin{tabularx}{\textwidth}{LLCC}
		\toprule
		\textbf{Error Model} & \textbf{Method} & $\boldsymbol{\langle}$\textbf{\emph{f}\textsubscript{\emph{relaxed}}}$\boldsymbol{\rangle}$   \textbf{(\%)}&  $\boldsymbol{\langle}$\textbf{\emph{f}\textsubscript{\emph{unrelaxed}}}$\boldsymbol{\rangle}$   \textbf{(\%)} \\ \midrule
		Gaussian & AD & \( 94.79 \pm 1.52 \) & \( 5.20 \pm 1.94 \) \\
		& Mclust & \( 95.76 \pm 1.72 \) & \( 3.93 \pm 1.74 \) \\ \midrule
		Student \( t \) & AD & \( 56.10 \pm 8.30 \) & \( 45.68 \pm 9.06 \) \\
		& Mclust & \( 69.75 \pm 5.05 \) & \( 31.50 \pm 8.46 \) \\ \toprule
	\end{tabularx}
\end{table}

\vspace{-9pt}
%%%%%%%%%%%%%%%%%%%%%%%%%%%%%%%%%%%%%%%%%%%%%%%%%%%%%%%%%%

\begin{table}[H]
	\centering
	\caption{Paired Wilcoxon signed-rank tests comparing Gaussian and Student-\(t\) error models for the cluster-by-cluster recovery fractions reported in Table~\ref{tab1}.}
	\label{wilcox}
	\begin{tabularx}{\textwidth}{LLLC}
		\toprule
		\textbf{State} & \textbf{Method} & \textbf{Comparison} & \textbf{\emph{p}-Value} \\
		\midrule

		Relaxed 
		& AD 
		& Gaussian vs. Student-\(t\) 
		& <10\textsuperscript{$-$70}  \\
		
		Relaxed 
		& Mclust 
		& Gaussian vs. Student-\(t\) 
		& <10\textsuperscript{$-$70}  \\
		
		Unrelaxed 
		& AD 
		& Gaussian vs. Student-\(t\) 
		& <10\textsuperscript{$-$200}\\
		
		Unrelaxed 
		& Mclust 
		& Gaussian vs. Student-\(t\) 
		& <10\textsuperscript{$-$200} \\
		
		\bottomrule
	\end{tabularx}
\end{table}

%%%%%%%%%%%%%%%%%%%%%%%%%%%%%%%%%%%%%%

In all cases, the~differences are highly significant, with~\(p<10^{-70}\) for relaxed systems and \(p<10^{-200}\) for unrelaxed systems. Therefore, the~changes discussed and observed in Table~\ref{tab1} are not merely descriptive fluctuations, but~reflect systematic differences induced by the adopted photometric error~model.

% =====================================
%

%\newpage

\subsection{A Spectroscopic Control Experiment with Mock Photometric~Redshifts}
\label{sec33}

{Before presenting the results of this control experiment, it is important to clarify its relationship to the analysis described in Sections~\ref{sec31} and \ref{sec32}. In~those sections, AD and Mclust were applied to velocity distributions constructed from the photometric redshifts available in the SDSS-based catalog. This procedure is useful because it reflects the practical situation in which only photometric information is available. However, it also has an important limitation: the observed photometric redshifts already contain uncertainties, possible systematic biases, selection effects, and~catastrophic outliers inherent to the photo-\(z\) estimation process. Therefore, when these redshifts are further resampled in Monte Carlo realizations, the~measured degradation reflects the combined effect of the adopted perturbation model and the intrinsic noise already present in the photometric catalog.}

{To isolate the impact of the error model itself, we performed a complementary sensitivity experiment in which synthetic photometric redshifts were generated directly from spectroscopic redshifts. The~purpose of this procedure is not to reproduce the full photo-\(z\) estimation pipeline from galaxy colors. Rather, spectroscopic redshift is used as a controlled reference redshift, and~well-defined Gaussian or Student-$t$ perturbations are applied to quantify how photometric-like uncertainties degrade the performance, and~consequently the classifications, of~the AD and Mclust diagnostics. In~this sense, the~experiment should be interpreted as a controlled sensitivity test of the dynamical diagnostics, rather than as a direct calibration of a specific photometric redshift catalog.}

For this test, we used an independent sample of clusters studied by~\cite{2018MNRAS.478.5473L}. Following their dynamical classification criteria, we considered a system to be relaxed when it simultaneously satisfies two conditions: a small projected offset between the brightest cluster galaxy and the X-ray peak, \(d_{\rm BCG-X}<0.01R_{500}\), and~a magnitude gap between the first and second brightest galaxies of \(\Delta m_{12}>1\). Systems not satisfying these criteria were classified as unrelaxed. After~applying these restrictions, the~final control sample consisted of 23 relaxed clusters and 26 unrelaxed~clusters.

{Although the \(\Gamma\) classification is based on optical morphology, the~criteria of~\cite{2018MNRAS.478.5473L}---BCG--X-ray offsets and \(\Delta m_{12}\)---are widely recognized in the literature as indicators of dynamical relaxation (e.g., \cite{2021A&A...655A.103Z,refId0}). The~use of the \(\Gamma\)-based sample is motivated by the need for an external dynamical reference that is independent of the velocity distribution being tested. Our goal is therefore complementary: we use \(\Gamma\) as a non-kinematic structural benchmark and then quantify how AD and Mclust classifications degrade when realistic photometric redshift error models are imposed. This sample thus provides an independent reference standard to validate the response of kinematic tests under controlled photometric errors.}

{Before introducing any simulated photometric redshift perturbation, we first evaluated the performance of AD and Mclust when applied directly to the velocity-offset distributions of the standard reference sample used \cite{2018MNRAS.478.5473L}.
In this spectroscopic baseline, AD correctly recovered 22 out of 23 relaxed systems ($95.65\%$), while Mclust recovered 21 out of 23 ($91.30\%$). For~the unrelaxed systems, AD recovered 25 out of 26 ($96.15\%$), and~Mclust recovered 24 out of 26 ($92.31\%$). These high recovery fractions (with performance comparable to that found by~\cite{2017AJ....154...96D} for Mclust and by~\cite{hou2009statistical}  for the AD test) demonstrate that, when applied to the original spectroscopic data, both diagnostics are intrinsically capable of identifying the reference dynamical state with excellent efficiency. Therefore, any degradation observed after the introduction of photometric uncertainties can be attributed to the effect of the assumed error models on the reconstructed velocity distributions, rather than to an intrinsic limitation of the tests themselves.}

In this experiment, the~simulated photometric redshift of each galaxy is constructed from its spectroscopic redshift according to
\begin{equation}
	z_{{\rm phot},i}^{\rm sim}
	=
	z_{{\rm spec},i}
	+
	(1+z_{{\rm spec},i})\epsilon_i,
	\label{eq:zphot_from_zspec}
\end{equation}

\noindent where \(\epsilon_i\) represents the normalized photometric error. For~comparison purposes, we consider the same two error models already used in previous sections for this experiment. In~the first, we assume a Gaussian distribution,
\begin{equation}
	\epsilon_i \sim \mathcal{N}(0,\sigma^2),
	\label{eq:gaussian_error_zspec}
\end{equation}

\noindent with \(\sigma = 0.02\). This case represents an idealized scenario in which photometric errors only broaden the redshift distribution, without~introducing heavy tails or extreme deviations, e.g.,~\cite{2019ApJ...884..164Y}. In~the second case, we assume that the errors follow a Student-\(t\) distribution with a scale adjusted to the typical size of the photometric error
\begin{equation}
	\epsilon_i \sim \sigma t_{\nu},
	\label{eq:tstudent_error_zspec}
\end{equation}

\noindent also with \(\sigma = 0.02\).{
For comparability purposes between error models (Gaussian and t with different \(\nu\)), we set \(\sigma=0.02\), a~value representative of the typical dispersion of photometric errors in surveys such as the SDSS for galaxies up to \(z\sim0.4\) (e.g., \cite{2006A&A...457..841I}). This simplification isolates the effect of the distribution shape (heavy tails vs. Gaussian) without the complexity of \(\sigma_{z,i}\) variation between galaxies}, but~allowing for heavier tails. We evaluated \(\nu = 2, 3\) and \(5\), such that \(\nu=2\) represents a more severe scenario with very heavy tails, \(\nu=5\) approaches a behavior closer to Gaussian, and~\(\nu=3\) provides an intermediate~case. 

The objective of this procedure is not to directly calibrate the actual bias of the observed photometric redshifts, but~rather to provide a model-dependent reference to quantify the expected degree of degradation in the recovery of AD and Mclust diagnostics under controlled photometric error prescriptions. Therefore, the~results obtained from \(z_{\rm spec} \rightarrow z_{\rm phot}^{\rm sim}\) should be interpreted as a sensitivity test, or~stress experiment, and~not as a direct calibration of the actual photometric catalog performance. Figure~\ref{lopes2018} and Table~\ref{control_samples} show the results obtained from this~experiment.

% ================================

% ================================
\vspace{-8pt}

\begin{figure}[H]
	\centering
	\includegraphics[width=1\linewidth]{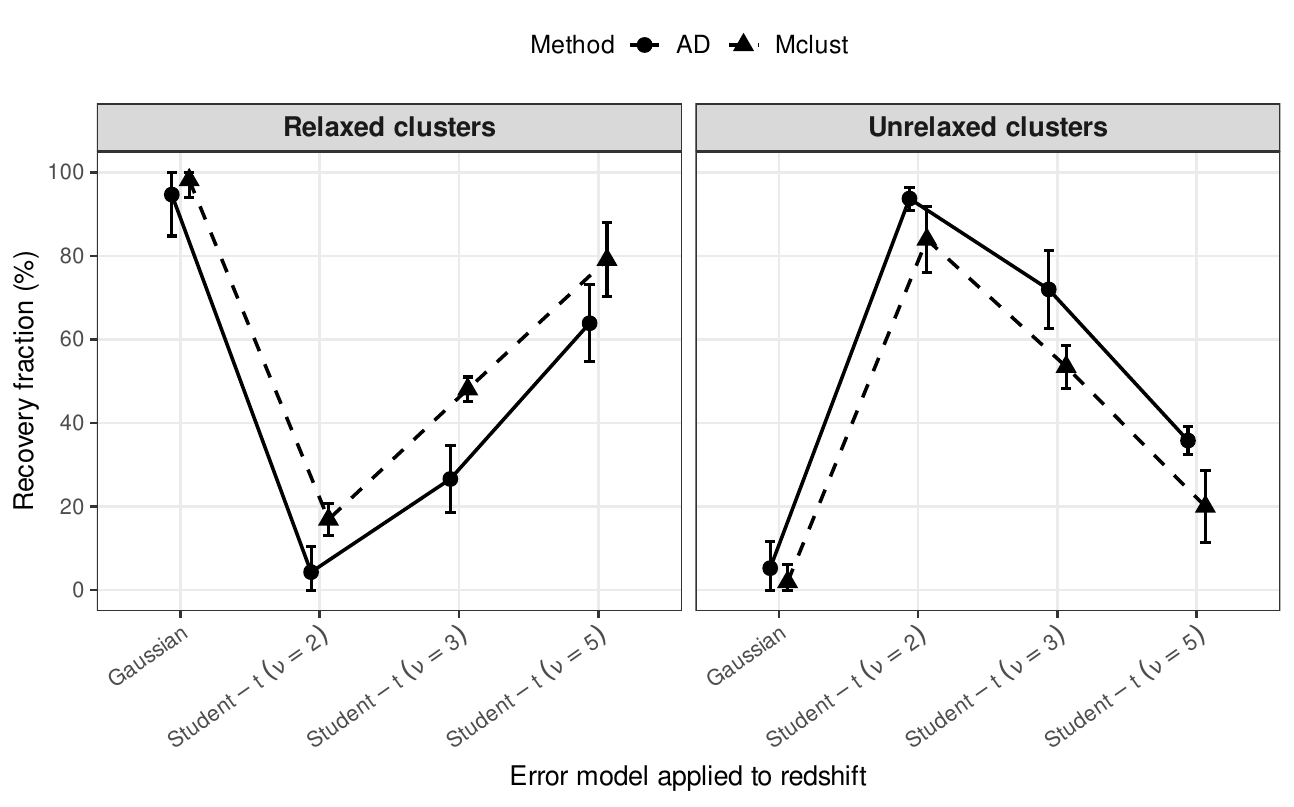}
	\caption{
		Average recovery fraction via AD and Mclust tests for relaxed (left) and unrelaxed (right) clusters applying Gaussian and Student-\(t\) error models to photometric redshifts generated from spectroscopic measurements. The points (filled circles - AD and filled triangle - Mclust) represent the percentage efficiency of the tests in correctly classifying each dynamic group for 1000 Monte Carlo realizations of the generated photometric sample. The error bars represent the standard deviation of the values ​​obtained for each of the 1000 realizations.
	}	
	\label{lopes2018}
\end{figure}

The left panel on Figure~\ref{lopes2018} corresponds to the relaxed sample, while the right panel corresponds to the unrelaxed sample. The~results show a strongly asymmetrical response of the AD and Mclust diagnostics to the different error~models.

%=====================================================

\begin{table}[H]
	\centering
	\caption{Recovery fractions for the relaxed and unrelaxed cluster samples in the spectroscopic control experiment. Mock photometric redshifts were generated from \(z_{\rm spec}\) assuming Gaussian and Student-\(t\) error~models.}
	\label{control_samples}
	\begin{tabularx}{\textwidth}{llCccc}
		\toprule
		\textbf{Sample} & \textbf{Method} & \textbf{Gaussian} & \textbf{\emph{t}, \emph{v} = 2} & \textbf{\emph{t}, \emph{v} = 3} & \textbf{\emph{t}, \emph{v} = 5} \\
		\midrule
		\multirow{2}{*}{Relaxed}
		& AD \( (\%) \)
		& \(94.7 \pm 9.9\) 
		& \(4.3 \pm 6.06\) 
		& \(26.6 \pm 7.98\) 
		& \(63.9 \pm 9.18\) \\
		& Mclust  \( (\%) \) 
		& \(98.2 \pm 4.1\) 
		& \(16.9 \pm 3.73\) 
		& \(48.1 \pm 2.92\) 
		& \(79.1 \pm 8.83\) \\
		\midrule
		\multirow{2}{*}{Unrelaxed}
		& AD  \( (\%) \)
		& \(5.25 \pm 6.37\) 
		& \(93.75 \pm 2.75\) 
		& \(72.0 \pm 9.36\) 
		& \(35.8 \pm 3.44\) \\
		& Mclust  \( (\%) \)
		& \(1.95 \pm 4.19\) 
		& \(83.95 \pm 7.91\) 
		& \(53.5 \pm 5.16\) 
		& \(20.0 \pm 8.68\) \\
		\bottomrule
	\end{tabularx}
\end{table}

%==============================================================

{Further, to~provide a more standard and comprehensive evaluation of our classification results, we consolidated the true/false positive and negative detection rates into standard machine learning metrics: accuracy, precision, recall, and~F1-score. Table~\ref{metrics} summarizes these metrics for both the AD and Mclust diagnostics across the different error models. Since the exact ratio of relaxed to unrelaxed clusters can vary depending on the specific survey and selection criteria, the~accuracy, precision, and~F1-score values were computed assuming a balanced baseline sample ($50\%$ relaxed, $50\%$ unrelaxed) to isolate the intrinsic performance of the tests.}

\begin{table}[H]
	\centering
	\caption{Classification metrics derived from the recovery fractions in Table~\ref{control_samples}. The~relaxed/Gaussian class is treated as the positive~class.}
	\label{metrics}
	\begin{tabularx}{\textwidth}{Llcccc}
		\toprule
		\textbf{Error Model} & \textbf{Method} & \textbf{Accuracy} & \textbf{Precision} & \textbf{Recall} & \textbf{F1-Score} \\
		\midrule
		Gaussian & AD & 50.0\% & 50.0\% & 94.7\% & 65.4\% \\
		Gaussian & Mclust & 50.1\% & 50.0\% & 98.2\% & 66.3\% \\
		\midrule
		Student-$t$ ($\nu=2$) & AD & 49.0\% & 40.8\% & 4.3\% & 7.8\% \\
		Student-$t$ ($\nu=2$) & Mclust & 50.4\% & 51.3\% & 16.9\% & 25.4\% \\
		Student-$t$ ($\nu=3$) & AD & 49.3\% & 48.7\% & 26.6\% & 34.4\% \\
		Student-$t$ ($\nu=3$) & Mclust & 50.8\% & 50.8\% & 48.1\% & 49.4\% \\
		Student-$t$ ($\nu=5$) & AD & 49.9\% & 49.9\% & 63.9\% & 56.0\% \\
		Student-$t$ ($\nu=5$) & Mclust & 49.6\% & 49.7\% & 79.1\% & 61.1\% \\
		\bottomrule
	\end{tabularx}
\end{table}

{The metrics in Table~\ref{metrics} quantitatively confirm the systematic patterns already identified in the recovery fractions discussed in Sections~\ref{sec31} and~\ref{sec32}. Under~Gaussian errors, the~high recall for relaxed clusters ($94.7\%$ for AD and $98.2\%$ for Mclust) corroborates the large $f_{\rm relaxed}$ values shown in Figure~\ref{gamma_relaxed}. However, the~low precision values ($\sim$50\%) indicate that this apparent success is accompanied by strong contamination of the relaxed class by intrinsically unrelaxed systems, as~also suggested by the poor recovery of non-Gaussian systems in Figure~\ref{gamma_unrelaxed}. For~the Student-$t$ error model, the~progressive decrease in the recall of relaxed clusters---from $\sim$64--$79\%$ for \(\nu=5\) to $\sim$4--$17\%$ for \(\nu=2\)---and the corresponding increase in the recovery of unrelaxed systems show that heavy-tailed errors alter the balance between the two dynamical classes. The~F1-scores summarize this trade-off: moderate values under Gaussian errors reflect the imbalance between high recall and low precision, whereas the sharp drop for \(\nu=2\), particularly for AD, indicates that strongly heavy-tailed errors severely compromise the identification of relaxed systems.}

{It is important to emphasize that the accuracy values remain close to \(50\%\) across most error models and methods. This should not be interpreted as evidence that the tests behave as random classifiers. Instead, similar accuracy values arise from two distinct systematic regimes. Gaussian errors preferentially produce Gaussian-like velocity distributions, thereby preserving the classification of relaxed systems while masking the non-Gaussian signatures of disturbed clusters. In~contrast, heavy-tailed Student-$t$ errors preferentially enhance non-Gaussian features, improving the recovery of unrelaxed systems but also producing artificial non-Gaussianity in intrinsically relaxed clusters. Thus, Table~\ref{metrics} shows that photometric redshift uncertainties introduce an asymmetric bias in the inferred dynamical state, rather than a simple uniform degradation of performance. Consequently, although~AD and Mclust remain useful for assessing the relative statistical impact of different photometric error models, their reliability as absolute dynamical classifiers depends critically on the adopted error distribution and should ideally be calibrated for the specific survey under consideration.}

\subsection{Dependence on Cluster~Richness}
\label{sec34}

The results presented in the previous sections demonstrate that the recovery of dynamical states depends strongly on the assumed photometric error model, with~Gaussian and Student-$t$ errors producing systematically different and asymmetric classification biases. However, these analyses were performed by averaging over the entire sample, potentially masking variations in performance that depend on cluster properties. In~particular, the~number of member galaxies---or richness---may play a critical role: richer clusters provide larger samples, reducing stochastic sampling variance and potentially improving the stability of the velocity distribution estimates (e.g., \cite{Ferragamo2021Velocity}). Conversely, larger samples may also accumulate more catastrophic outliers, which could amplify the distortions induced by heavy-tailed errors. Understanding how richness modulates the effects of photometric errors is therefore essential for interpreting the practical applicability of these diagnostics, and~for assessing whether the bias patterns identified in Sections~\ref{sec31}--\ref{sec33} are universal or depend on cluster observables. To~address this question, we now investigate whether the recovery fraction correlates with cluster~richness.

In this section, we investigate whether the recovery fraction correlates with cluster richness, considering both dynamical states (relaxed and unrelaxed). To~do this, we use Beta regression, a~statistical model developed specifically for continuous response variables that take values in the interval $(0,1)$, such as rates, proportions, and~percentages~\cite{Ferrari01082004, 1471082X03st053oa}. 

We assumed that the observed fraction $y_i$ for cluster $i$ follows a Beta distribution parameterized in terms of its mean $\mu_i$ and precision parameter $\phi$,
\begin{equation}
	y_i \sim \mathrm{Beta}(\mu_i \phi,\,(1-\mu_i)\phi),
\end{equation}
\noindent where $\mathbb{E}(y_i)=\mu_i$ and
\begin{equation}
	\mathrm{Var}(y_i)=\frac{\mu_i(1-\mu_i)}{1+\phi}.
\end{equation}

The mean was modeled through a logit link as
\begin{equation}
	\text{logit}(\mu_i)=\beta_0+\beta_1\,n_{\mathrm{obj},i},
\end{equation}
\noindent where $n_{\mathrm{obj},i}$ is the number of galaxies within $r_{500}$, 
and the linear dependence in logit was considered as a first-order approximation. This choice is motivated by the absence of a priori constraints on the functional form of this relationship and allows for a straightforward interpretation of the marginal effect of richness on recovery performance. 
Furthermore, in~the baseline formulation, the~precision parameter $\phi$ was assumed to be constant across observations, i.e.,~$\phi_i = \phi$. Since our primary interest lies in the mean recovery behavior rather than heteroscedasticity, we adopt this choice for all~clusters.

Statistical significance of the slope $\beta_1$ was evaluated via Wald $z$-tests, and~95\% confidence intervals were obtained from the asymptotic normality of the maximum likelihood estimators. Average marginal effects (AMEs) were computed as the sample average of 
$\partial \mu_i / \partial n_{\mathrm{obj}} = \beta_1 \, \mu_i (1 - \mu_i)$ 
to aid physical interpretation. All analyses were performed in \texttt{R} using the \texttt{betareg} package \citep{JSSv048i11}. Figure~\ref{numeric} and Table~\ref{beta_results1} show the results~obtained.

\begin{figure}[H]
	
	\hspace{-3pt}\includegraphics[width=1.0\linewidth]{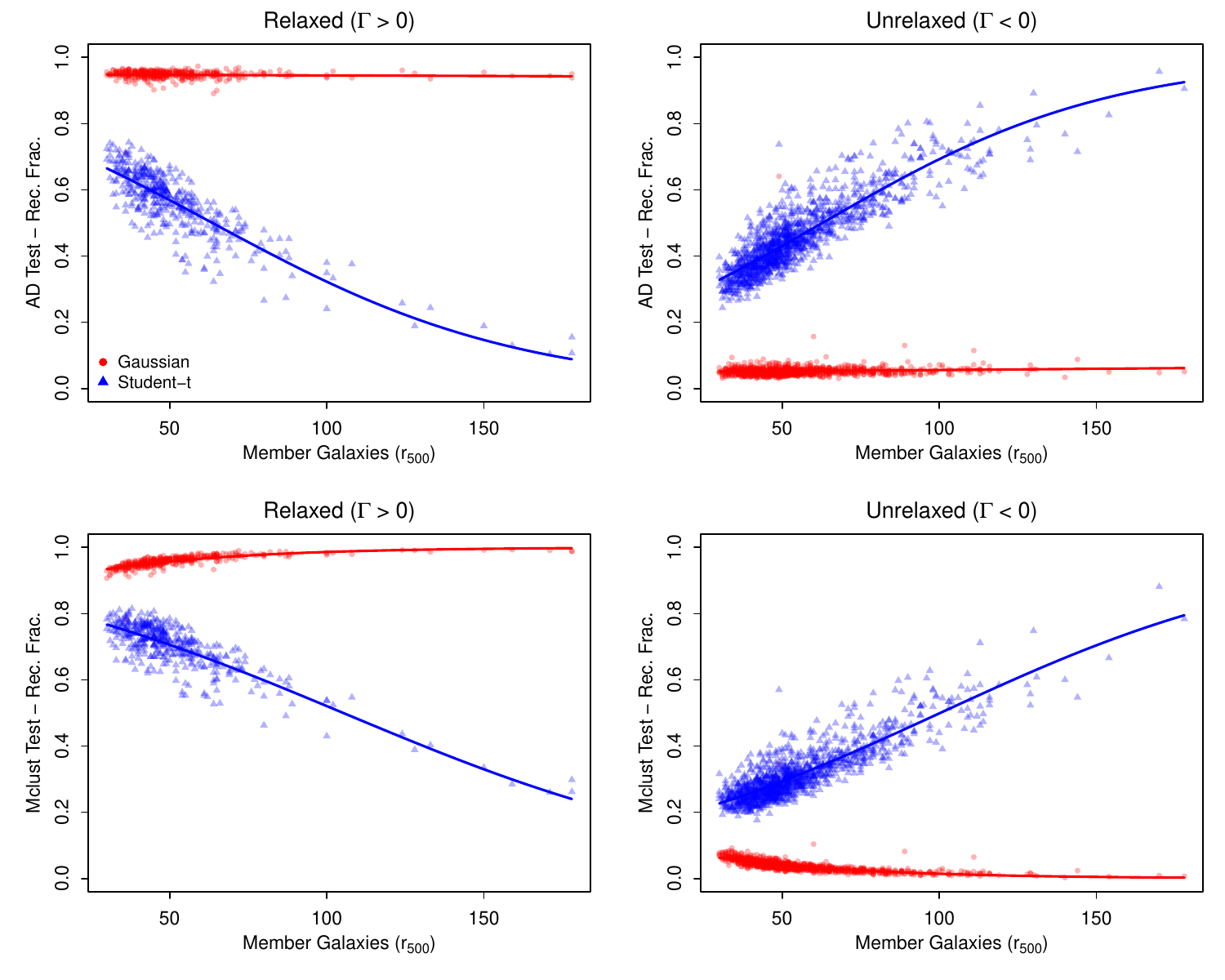}
	\caption{Recovery fraction (Rec. Frac.) of normality tests---AD (top panels) and Mclust (bottom panels)---as a function of the number of member galaxies within $r_{500}$. Results are shown separately for relaxed ($\Gamma > 0$, left) and unrelaxed ($\Gamma < 0$, right) clusters. In~each panel, solid curves represent beta regression fits for Gaussian (red) and Student-$t$ (blue) underlying distributions. The~recovery fraction indicates the proportion of Monte Carlo realizations in which the true model was correctly~identified.}	
	\label{numeric}
\end{figure}

\begin{table}[H]
	
	\footnotesize
	\setlength{\tabcolsep}{8pt}
	
	\caption{Beta regression results for the dependence of the recovery fraction on cluster richness. Average marginal effects (AMEs) are reported on the response scale, with~95\% confidence intervals. }%
	\label{beta_results1}
	\begin{tabularx}{\textwidth}{lcCccc}
		\toprule
		\textbf{Test} & \textbf{Dist.} & \textbf{AME} ($\boldsymbol{\times}$\textbf{10\textsuperscript{$-$4}}) & \textbf{\emph{p}-Value} & \textbf{Pseudo-R\textsuperscript{2}} \\
		\midrule
		AD/Relaxed & Gauss 
		& $-0.296\,[-0.852,\,0.260]$ 
		& 0.30 & 0.003 \\
		
		& $t$-Stud 
		& $-50.2\,[-53.5,\,-46.9]$ 
		& <$10^{-15}$ & 0.738 \\
		
		AD/Unrelaxed & Gauss 
		& $+0.679\,[0.329,\,1.029]$ 
		& $1.5\times10^{-4}$ & 0.016 \\
		
		& $t$-Stud 
		& $+54.2\,[52.5,\,55.9]$ 
		& <$10^{-15}$ & 0.772 \\
		
		Mclust/Relaxed & Gauss 
		& $+9.48\,[8.77,\,10.19]$ 
		& <$10^{-15}$ & 0.644 \\
		
		& $t$-Stud 
		& $-33.4\,[-35.4,\,-31.4]$ 
		& <$10^{-15}$ & 0.711 \\
		
		Mclust / Unrelaxed & Gauss 
		& $-8.28\,[-8.64,\,-7.92]$ 
		& <$10^{-15}$ & 0.694 \\
		
		& $t$-Stud 
		& $+37.7\,[36.3,\,39.1]$ 
		& <$10^{-15}$ & 0.749 \\
		\bottomrule
	\end{tabularx}

	\footnotesize
	\textit{Note}: AME = average marginal effect, computed as the sample average of the marginal effects on the response scale. Confidence intervals are derived from asymptotic normality.
\end{table}

Based on the results shown in Figure~\ref{numeric} and Table~\ref{beta_results1}, we observe method-dependent behaviors under Gaussian error assumptions: the AD test shows negligible richness dependence for relaxed clusters (AME $-0.296 \times 10^{-4}$, $p = 0.30$, Pseudo-$R^2 = 0.003$) and only weakly significant dependence for unrelaxed clusters ($+0.679 \times 10^{-4}$, $p = 1.5 \times 10^{-4}$, Pseudo-$R^2 = 0.016$). In~contrast, Mclust exhibits strong but opposing correlations — positive for relaxed ($+9.48 \times 10^{-4}$, $p < 10^{-15}$, Pseudo-$R^2 = 0.644$) and negative for unrelaxed ($-8.28 \times 10^{-4}$, $p < 10^{-15}$, Pseudo-$R^2 = 0.694$) clusters. These divergent Gaussian-model patterns reflect fundamental methodological differences: AD assesses goodness-of-fit to a reference distribution, while Mclust performs model selection among competing parametric forms, leading to opposite richness dependencies, particularly under Gaussian error~assumptions.

In contrast, the~Student-$t$ model reveals strong, consistent, state-dependent correlations across both tests with substantially larger effect sizes. For~relaxed clusters, richer systems exhibit \emph{lower} recovery fractions. This indicates that larger galaxy samples increase the impact of catastrophic outliers, which distort the velocity distribution and reduce the apparent Gaussianity (AD: $-50.2 \times 10^{-4}$; Mclust: $-33.4 \times 10^{-4}$; both $p < 10^{-15}$, Pseudo-$R^2 \approx 0.71$--$0.74$). Conversely, for~unrelaxed clusters, richer systems show \emph{higher} recovery fractions, indicating that larger samples better constrain the intrinsically complex velocity distributions characteristic of perturbed systems (AD: $+54.2 \times 10^{-4}$; Mclust: $+37.7 \times 10^{-4}$; both $p < 10^{-15}$, Pseudo-$R^2 \approx 0.75$--$0.77$).

Importantly, the~Student-$t$ marginal effects are 10--100-times larger than their Gaussian counterparts, particularly in the AD case, indicating a substantially improved fit relative to the null model (Pseudo-$R^2 > 0.7$) and demonstrating that outlier-robust error modeling can significantly alter the richness--recovery relationship. This pronounced asymmetry implies that photometric dynamical classification does not benefit uniformly from increased sample size, instead suggesting that the impact of richness depends on both the dynamical state and the statistical framework~adopted.

This opposite behavior reinforces the interpretation that photometric noise affects the two dynamical states through different mechanisms: while relaxed systems are increasingly perturbed by the accumulation of outliers, unrelaxed systems benefit from improved sampling of their intrinsic complexity. As~a result, the~asymmetry in the recoverability of dynamical states is not only preserved but becomes more pronounced as a function of~richness.

%%%%%%%%%%%%%%%%%%%%%%%%%%%%%

\section{Discussion}

The results presented highlight a fundamental limitation in the use of photometric redshifts for dynamical studies of galaxy clusters. While simple statistical diagnostics such as AD and Mclust remain attractive due to their low computational cost and ease of application, our analysis shows that their performance is intrinsically asymmetric: relaxed systems can still be identified with moderate reliability, whereas unrelaxed systems are systematically harder to~detect. This asymmetric bias parallels \cite{2025Univ...11...82C}, who found that limited depth (magnitude cuts) also underestimates substructures, showing that both depth and redshift quality bias dynamical classifications.

The results of our richness analysis (Section~\ref{sec34}) suggest that the increased number of member galaxies expected in future deep photometric surveys such as LSST may improve the recovery of dynamically disturbed systems under heavy-tailed photometric error models.
Specifically, for~unrelaxed clusters ($\Gamma < 0$), we find that richer systems yield higher $f_{\rm unrelaxed}$ fractions when errors are modeled with Student-$t$ distributions. This improvement arises because larger samples provide sufficient statistical power to recover the intrinsically complex velocity distributions---such as multimodality or heavy tails---that are characteristic of merging or substructured systems, even in the presence of photometric uncertainties. Conversely, for~relaxed clusters, increasing richness under Student-$t$ errors leads to slightly lower recovery fractions, since the accumulation of catastrophic outliers can artificially distort an otherwise Gaussian distribution. This asymmetry implies that photometric dynamical classification does not benefit uniformly from increased sample size. For~massive systems up to $z \sim 0.5$, one can expect typical memberships of $N_{500} \sim 100$--$200$ (e.g., \cite{Capozzi2011, 2018MNRAS4734077P}), suggesting that the improved sampling may enhance the identification of disturbed systems compared to the SDSS-era samples analyzed~here.

However, this potential gain must be weighed against a competing effect: the degradation of photometric redshift quality with increasing redshift. Beyond~$z \gtrsim 0.5$, photo-$z$ errors become intrinsically more difficult to calibrate, and~the fraction of catastrophic outliers can increase significantly in the absence of representative spectroscopic training samples (e.g., \cite{2008ApJ...689..709O, 2025RAA25e5021L, 2025ApJ...983..173D}). Recent studies suggest that poorly matched training sets can amplify the outlier fraction by up to a factor of four, directly affecting the reliability of velocity-based diagnostics \cite{2024ApJ...967L...6M}.

These two effects---increased richness and increased photo-$z$ degradation at high redshift---act in opposite directions. The~net impact on the recovery of dynamical states will therefore depend on the redshift distribution of the cluster sample and on the specific photo-$z$ calibration strategy adopted. For~low-to-moderate redshift clusters ($z \lesssim 0.5$), where photo-$z$ quality remains relatively high, the~richness gain may dominate, enabling improved identification of disturbed systems compared to SDSS-era samples. For~high-redshift clusters ($z \gtrsim 0.5$), the~degradation in photo-$z$ quality may outweigh the benefits of larger memberships, potentially leading to performance similar to or worse than what we have quantified here. This is because the intrinsic velocity dispersion of massive clusters, typically a few $10^2$--$10^3 \, {\rm km \cdot s^{-1}}$, corresponds to a redshift imprint $\sigma_z \simeq (1+z_{\rm cl})(\sigma_v/c)$ that remains much narrower than typical photometric redshift uncertainties, especially when catastrophic outliers become more frequent. As~a result, the~velocity structure associated with cluster dynamics becomes increasingly difficult to recover from photometric redshifts alone, and~the separation between kinematic components---such as those arising from mergers or substructures---becomes progressively blurred, reducing the detectability of non-Gaussian~features.

It is also important to emphasize that our goal is not to dismiss the use of photometric data for dynamical studies, but~rather to establish its operational limits. In~this context, ongoing efforts within the community are highly relevant. Forecasting approaches based on Fisher matrix formalism, for~example (e.g., \cite{2006ApJ...636...21M, 2018arXiv180901669T}), illustrate the level of detail required to incorporate photometric uncertainties in next-generation surveys. Within~this framework, however, photometric redshift errors are typically described through simplified parametric models---such as bias, scatter, and~the fraction of catastrophic outliers---indicating that their statistical treatment remains an evolving challenge. The~present work follows a complementary direction by explicitly exploring how different error distributions, including Student-$t$ models, impact the recovery of dynamical~information.

A natural next step is to apply the same methodology to realistic LSST-like mock catalogs. This would move the analysis beyond the idealized Gaussian and Student-$t$ prescriptions adopted here and allow one to assess how survey depth, selection effects, photo-$z$ calibration, galaxy population evolution, and~spatially varying observational uncertainties jointly affect the recovery of relaxed and unrelaxed systems. Mock catalogs tailored to LSST, such as those based on the DESC DC2 simulation \cite{2021ApJS..253...31L}, provide a particularly promising framework, as~they incorporate realistic photometry, survey-depth variations, and~complex galaxy population evolution. Applying our methodology to such catalogs would allow a direct evaluation of whether the limitations identified here persist, weaken, or~become more severe under LSST-like observing~conditions.

Our work therefore provides a cautionary baseline: even before the full complexity of LSST-like observations is included, photometric redshift uncertainties already impose a strong asymmetric limitation on dynamical classification. Our analysis further suggests that future photometric dynamical studies should prioritize: (i) robust photo-$z$ calibration, particularly at high redshift; (ii) explicit modeling of catastrophic outliers; and (iii) validation with realistic survey mock catalogs that incorporate both richness and redshift-dependent error distributions. Without~substantial improvements in photo-$z$ calibration and catastrophic-outlier mitigation, reliable dynamical classification based primarily on photometric redshifts will remain challenging, particularly for the identification of dynamically disturbed~systems.

%%%%%%%%%%%%%%%%%%%%%%%%%%%%%%%%%%%%%%%%%%%%%%%%%
%%%%%%%%%%%%%%%%%%%%%%%%%%%%%%%%%%%%%%%%%%%%%%%%%

\authorcontributions{
Conceptualization, A.P.C.; methodology, A.P.C. and A.L.B.R.; software, A.P.C.; validation, A.P.C. and A.L.B.R.; formal analysis, A.P.C.; investigation, A.P.C., A.L.B.R., Z.L.W and F.R.M.-N.; data curation Z.L.W and A.P.C; writing--original draft preparation, A.P.C.; writing--review and editing, A.P.C. and A.L.B.R. All authors have read and agreed to the published version of the manuscript.}

\funding{This work was supported by the Coordenação de Aperfeiçoamento de Pessoal de Nível Superior (CAPES).}

\dataavailability{We used data from SDSS through the study by~\cite{2013MNRAS.436..275W}.} 
%=====================================================

\acknowledgments{We are deeply grateful to the anonymous reviewers for their insightful comments and constructive suggestions, which significantly improved the quality of this manuscript.}

\conflictsofinterest{The 
authors declare no conflicts of interest.}

%%%%%%%%%%%%%%%%%%%%%%%%%%%%%%%%%%%%%%%%%%
\begin{adjustwidth}{-\extralength}{0cm}

\reftitle{References}


%%%%%%%%%%%%%%%%%%%%%%%%%%%%%%%%%%%%%%%%%%
\PublishersNote{}
\end{adjustwidth}
\end{document}